\documentclass[fleqn,usenatbib]{mnras}

\usepackage{amsmath}
\usepackage{mathptmx}
\usepackage{txfonts}
\usepackage{graphicx}
\usepackage{adjustbox}
\usepackage{caption}
\usepackage{subcaption}
\usepackage[T1]{fontenc}

\DeclareRobustCommand{\VAN}[3]{#2}
\let\VANthebibliography\thebibliography
\def\thebibliography{\DeclareRobustCommand{\VAN}[3]{##3}\VANthebibliography}

\usepackage{graphicx}	% Including figure files
\usepackage{amsmath}	% Advanced maths commands
\title[Spectral broadening and Turbulence]{A turbulence index independent framework for deriving solar wind speed and coronal electron density from radio spectral broadening}

\author[Aggarwal et al.]{
Keshav Aggarwal$^{1}$\thanks{E-mail: keshavagg1098@gmail.com (KA)},
R.\ K.\ Choudhary$^{2}$,
Abhirup Datta$^{1}$,
Soumyaneal Banerjee$^{2,3}$,
Takeshi Imamura$^{4}$,
H. Ando$^{5}$\\
$^{1}$Department of Astronomy, Astrophysics and Space Engineering (DAASE), Indian Institute of Technology Indore, Indore, Madhya Pradesh 453552, India\\
$^{2}$Space Physics Laboratory (SPL), Indian Space Research Organization, Vikram Sarabhai Space Centre, Thiruvananthapuram, Kerala 695022, India\\
$^{3}$Research Centre, University of Kerala, Thiruvananthapuram 695034, India\\
$^{4}$Graduate School of Frontier Sciences, The University of Tokyo, Kashiwa, Japan\\
$^{5}$Faculty of Science, Kyoto Sangyo University, Kyoto, Japan
}

\date{Accepted XXX. Received 2025 May 6; in original form ZZZ}

\pubyear{\the\year{}}

\begin{document}
\label{firstpage}
\pagerange{\pageref{firstpage}-\pageref{lastpage}}
\maketitle

\begin{abstract}
We present a turbulence index independent framework for simultaneously deriving solar wind velocity and coronal electron density in the near-Sun region using the spectral broadening of spacecraft radio signals. The formulation accommodates arbitrary turbulence spectral indices ($p$), providing a direct analytical link between the observed Doppler spectra and underlying plasma parameters without assuming a fixed turbulence regime. This generalization extends conventional radio occultation techniques and enables consistent interpretation across multiple radio frequencies. We apply the method to X-band ($\sim$ 8.41 GHz) radio occultation measurements from JAXA's Akatsuki spacecraft during the 2016 and 2022 Venus - Earth superior conjunctions, spanning heliocentric distances of 1.4 - 10 $R_{\odot}$ and sampling both equatorial streamer regions and mid-latitude coronal holes. The retrieved electron densities exhibit systematic trends consistent with empirical coronal models and in-situ observations. By coupling the measured spectral widths with a turbulence-based frequency-scaling relation, we obtain a compact expression that links spectral broadening, solar wind speed, and electron density, applicable for any turbulence index $p$. Fast-solar-wind intervals, characterized by nearly isotropic turbulence, yield speed estimates in close agreement with expectations, while the anisotropic nature of the slow solar wind introduces small but systematic deviations. Our results refine earlier work and demonstrate that explicit consideration of near-coronal turbulence anisotropy is essential for accurate solar-wind parameter retrievals.
\end{abstract}


\section{Introduction}

The solar wind is a continuous outflow of magnetized plasma from the solar corona, composed primarily of protons and electrons, with a small fraction of heavy ions \citep{Marsch1999, SchwennBook1990}. Its variability from steady fast and slow streams to transient events such as coronal mass ejections (CMEs) and solar energetic particle (SEP) bursts governs the space weather environment throughout the solar system \citep{Kilpua2016}. The solar atmosphere from which these winds originate is stratified into the photosphere, chromosphere, transition region, and corona. The photosphere, with a temperature of about 5800 K, forms the visible surface of the Sun. Above it, the temperature rises sharply through the chromosphere and transition region, reaching more than $10^6$ K in the outermost layer, the corona \citep{Marsch2006}. It is from this region that the solar wind originates and fills the entire heliosphere, with its properties strongly dependent on the magnetic and plasma conditions of the corona \citep{Kilpua2016}. 

Because the large-scale magnetic topology of the corona governs where plasma can expand outward, the source regions of different wind components follow directly from this structure. The fast wind originates from open magnetic field regions such as coronal holes \citep{Zirker1977}, while the slow wind is associated with helmet streamers and the vicinity of active regions \citep{Habbal_1997}. Transient enhancements from CMEs add further complexity. The solar wind emerges subsonically from the low corona and accelerates to supersonic and super-Alfv\'enic speeds beyond about 20-30 $R_{\odot}$ \citep{Sheeley1997, Kasper2016}. Mapping its speed and density profiles in the near-Sun region is essential to understanding its origin and the mechanisms that drive its acceleration.

The dominant candidate mechanism for supplying this acceleration is turbulent energy transfer. Alfv\'enic fluctuations generated in the lower atmosphere propagate upward, interact nonlinearly, and cascade to smaller spatial scales, where they dissipate and heat the plasma \citep{Tu1995, Cranmer2015, Shoda2019}. These turbulent interactions are believed to power both coronal heating and solar wind acceleration, leaving measurable signatures in the fluctuations of electron density and plasma velocity. Quantifying these fluctuations provides a direct means to diagnose the energy cascade and the physical conditions of the coronal plasma.

While turbulence has been widely recognized as a key driver of coronal heating, its detailed properties close to the Sun remain poorly constrained \citep{Tu1995, Cranmer2015, Shoda2019}. Classical turbulence theory predicts a power-law cascade with Kolmogorov scaling \citep{Kolmogorov1941, Woo1976, Armand1987}, but spacecraft and radio scintillation observations show systematic deviations from this ideal form. The spectral index of density fluctuations varies with heliocentric distance and latitude, reflecting the evolving anisotropy, intermittency, and the combined influence of Alfv\'enic and magnetosonic wave interactions \citep{Efimov2005, Efimov2008, Chashei2007}. Characterizing this transition from the structured inner corona to the freely expanding solar wind is critical for understanding energy transport, dissipation, and the development of turbulence in the heliosphere \citep{Goldreich1995, Shoda2018}.

Interpreting these deviations requires specifying the form of the underlying density fluctuation spectrum. The spatial spectrum of coronal density fluctuations is generally modeled as a three-dimensional, nearly isotropic power law, $\Phi_N(q) \propto q^{-p}$ \citep{Armand1987, Tatarskii1971}. While it has been established that this is not the case in the inner and middle corona and anisotropy is observed in this region, recent studies using data from the Aditya-L1 mission also have demonstrated that while slow solar wind shows anisotropic behavior, fast solar winds show isotropic behaviour \citep{Armstrong1990, Yadav2025}. In the inertial subrange of fully developed turbulence, energy cascades from large to small scales following a Kolmogorov slope of $p \approx 11/3$ \citep{Kolmogorov1941, Woo1976}. At smaller scales, beyond the inner dissipation limit, the spectrum steepens as energy is converted into heat. For radio waves propagating through such a medium, the “frozen-in” approximation \citep{Armand1987} assumes that plasma irregularities are advected with the bulk solar wind flow, producing measurable phase and frequency fluctuations in the received signal.

Radio occultation (RO) provides a powerful means of probing this turbulent medium. When spacecraft radio signals pass close to the Sun during conjunction, they traverse regions of strong electron density fluctuations, resulting in measurable changes in signal amplitude, phase, frequency, and polarization. Each of these quantities carries information about plasma density, velocity, and the turbulence spectrum along the line of sight. Early radio science experiments with the Mariner, Pioneer, Helios, and Viking missions demonstrated the diagnostic power of this technique \citep{Woo1976, Bird1982}. More recent measurements from Ulysses, Galileo, Mars Express, MESSENGER, and Akatsuki have extended this approach to different heliocentric latitudes and solar cycle conditions \citep{Patzold1996, Imamura2005, Wexler2019}. 

The key geometric parameter in radio occultation is the solar offset distance ($r$) which determines how deeply the ray penetrates into the corona. While in-situ probes such as the Parker Solar Probe (PSP) have approached within $\sim$13 $R_{\odot}$ \citep{Badman2023}, radio occultation can probe even closer, reaching down to $\sim$1.4 $R_{\odot}$, a region critical for understanding solar wind acceleration and turbulence onset. By analyzing the spectral broadening of spacecraft signals, RO enables remote estimation of electron density ($n_e$), solar wind velocity ($v$), and the spectral characteristics of turbulence in the near-Sun corona.

Building on this theoretical foundation, radio occultation measurements can be directly connected to physical parameters of the solar wind, including electron density ($N_e$) and radial velocity ($v$). As established in earlier studies, by analyzing Doppler spectral broadening in spacecraft signals across multiple frequencies, it is possible to derive these quantities \citep{Aggarwal2025a, Aggarwal2025b}. This unified approach enabled comparisons across heliocentric distances, latitudes, and different solar activity conditions, providing a means to study plasma structure, and solar wind acceleration, assuming a fixed turbulence spectrum of $p = 11/3$. This motivates a formulation that does not rely on a fixed spectral exponent. Here in this study, we have modified this approach to be able to be applicable for different cases without having to assume a fixed turbulence spectrum of $p = 11/3$, which is the value for a fully developed, isotropic Kolmogorov turbulence spectrum. In the following sections, we describe the methodology used to extract spectral indices from radio occultation data from the Japanese Akatsuki spacecraft and to quantify electron density and velocity of the solar wind in the near-Sun coronal region for varying heliocentric distances.

\begin{figure*}
    \centering
    \includegraphics[width=0.75\linewidth]{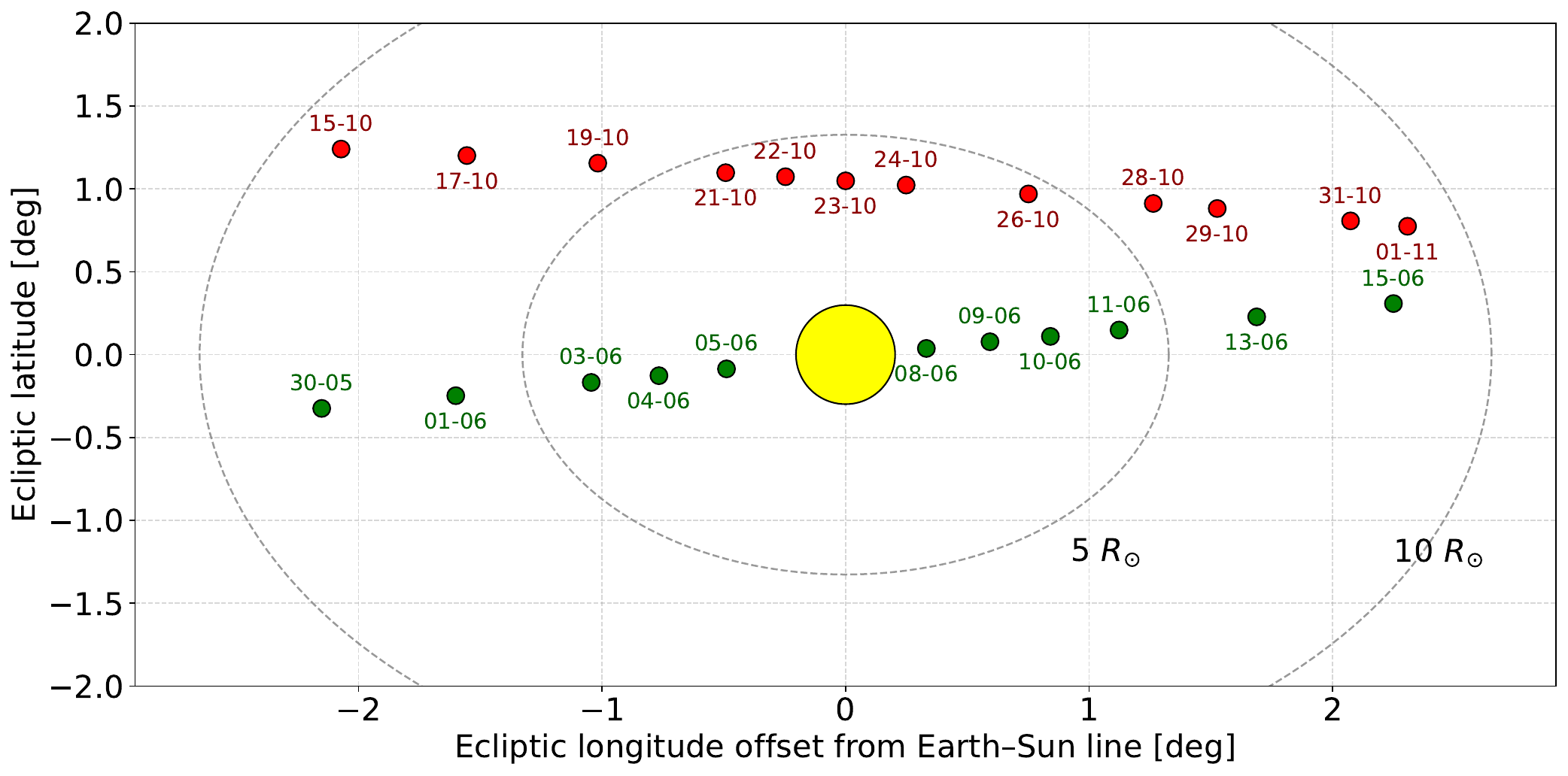}
    \caption{Geometry of the 2016 and 2022  radio occultation experiment shown. The points in green mark the position of the satellite during the 2016 occultation experiment, while the points in red mark the position during the 2022 occultation experiment.}
    \label{fig:schematic}
\end{figure*}

\section{Methodology}
\label{sec:method_data_2022}

X-band (8.41 GHz) radio occultation observations were carried out using the Venus Climate Orbiter (Akatsuki/VCO), which has been conducting regular solar conjunction experiments since entering orbit around Venus \citep{Chiba2022, Chiba2023, Aggarwal2025b}. In this study, we analyzed two observing campaigns: one in June 2016, during the declining phase of Solar Cycle 24, and another in October 2022, corresponding to the early rise of Solar Cycle 25. Together, these two epochs provide a unique opportunity to examine the near-Sun plasma under contrasting heliophysical conditions.

The spacecraft was equipped with a 1.6 m radial line slot antenna and an ultrastable oscillator (USO) of exceptional frequency stability, characterized by an Allan deviation of $<10^{-12}$  for the averaging time of $1-1000 $s. This configuration ensured highly coherent downlinks at X-band, which were received by the 64 m antenna of the Usuda Deep Space Center in Japan \citep{Oshima2011, Imamura2017}. The USO-induced frequency instability, $\sim8.4\times10^{-3}$ Hz at X-band, was negligible compared to the plasma-induced Doppler fluctuations \citep{Tripathi2022b}. The open-loop radio science data, recorded in CCSDS-RDEF format, were processed using established pipelines described in \citet{Tripathi2022, Aggarwal2025b}.

The 3.4$^\circ$ inclination of Venus’s orbit relative to the ecliptic allowed the radio ray paths to probe a wide range of heliographic latitudes while maintaining excellent signal-to-noise ratios. The geometry of Akatsuki’s orbit during the 2016 and 2022 conjunctions enabled line-of-sight (LOS) sampling of the near-Sun plasma across diverse heliocentric distances and latitudes. Figure \ref{fig:schematic} illustrates the geometry of these two observing campaigns. The green points mark the spacecraft positions during the 2016 occultation experiment, while the red points correspond to those of 2022. The 2016 observations sampled the trans-equatorial region of the corona during a period of low solar activity, whereas the 2022 campaign probed a coronal-hole region under moderately active conditions. This configuration provided an excellent opportunity to assess the influence of large-scale solar wind structures on radio signal propagation and spectral broadening.

\begin{figure*}
    \centering
    \includegraphics[width=0.65\linewidth]{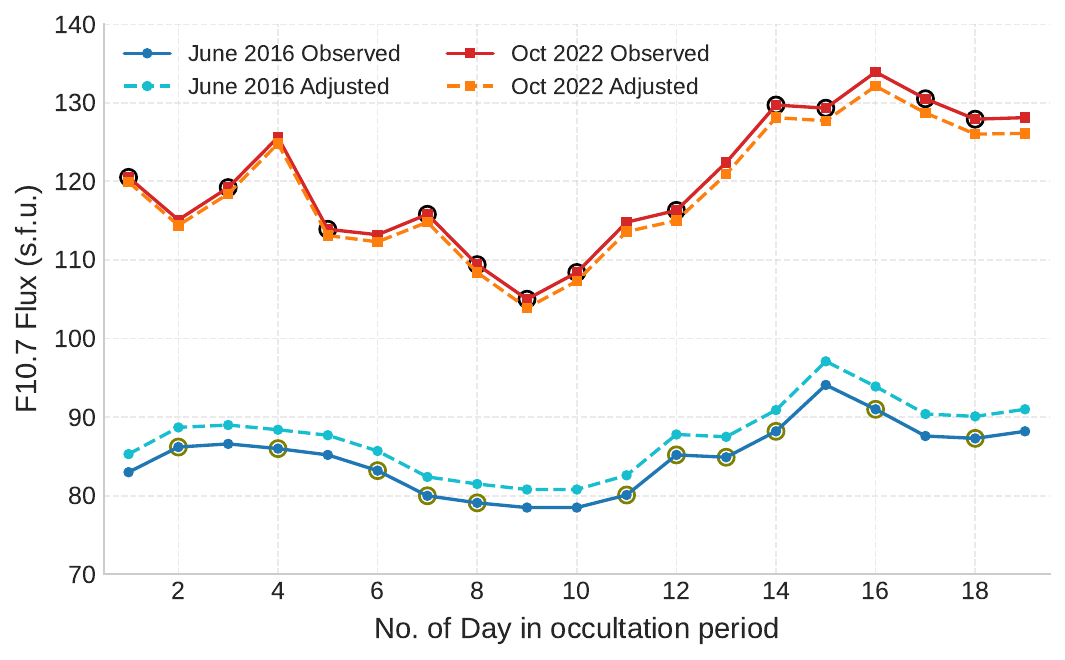}
    \caption{A comparison of the f10.7 values for the days of occultation experiments conducted in 2016 and 2022.}
    \label{fig:f107}
\end{figure*}

The contrast in solar activity between the two epochs is clearly evident in Figure \ref{fig:f107}. The blue solid and dashed curves represent the observed and adjusted F10.7 cm solar radio flux for 2016, while the red and orange curves show the corresponding values for 2022. Olive and black circles indicate the specific days of the occultation experiments. These two sets of observations, obtained under distinctly different solar wind regimes, one dominated by magnetically confined, compressive slow winds in the trans-equatorial region in 2016, and the other by Alf\'venic, low-compressibility fast wind in a coronal hole region in 2022 \citep{Jain_2024}, allow us to investigate the evolution of the coronal turbulence spectrum across contrasting plasma environments.

The raw CCSDS-RDEF data files were first converted into 1 s sampled carrier-frequency time series using the standard processing pipeline outlined in \citet{Aggarwal2025a}. This procedure transforms the voltage waveforms recorded by the ground station into a sequence of frequency measurements that represent the Doppler shifts imparted by the solar corona. Each one-second segment of this series was Fourier-transformed to obtain its power spectrum, to which a Gaussian profile was fitted. From the fitted parameters, three primary quantities were extracted: the total spectral power, the mean Doppler shift, and the spectral width ($B$) \citep{Woodman1985}.

The spectral width $B$, quantifies the extent to which the received radio signal is dispersed in frequency due to small-scale fluctuations in the refractive index of the turbulent coronal plasma. Physically, a larger $B$ indicates stronger plasma density irregularities and, consequently, enhanced turbulence and flow variations along the propagation path. The statistical uncertainty in $B$ was estimated as the standard deviation of repeated measurements, corresponding to approximately 15\% of its mean value, which dominates the overall error budget \citep{Aggarwal2025a, Aggarwal2025b}. The contribution from the spacecraft’s ultrastable oscillator remains negligible (Allan deviation $\sim10^{-12}$ Hz/Hz) \citep{Tripathi2023b}.

\subsection{From Spectral Broadening to solar wind Parameters}

The main goal of this study is to use the measured spectral broadening to infer meaningful physical quantities: the electron density in the corona ($N_e$) and the radial velocity of the solar wind ($v$). We assumed that the turbulence in the corona follows a Kolmogorov-type spectrum, which is a well-established description of how energy cascades from large to small scales in turbulent fluids. Using this assumption, and as established in \cite{Aggarwal2025b}, the electron density and radial solar wind velocity using a S-band signal can be calculated using the following empirical formulas:
\begin{equation} 
N_e = \frac{f}{r^2 \times R_{SP} \times r_{\odot} \times (1+ R_{SP})} \times \left( \frac{B_s}{c_0} \right)^{5/6} 
\label{ogden}
\end{equation} 
\begin{equation} 
v = k_0 \left[ \frac{r\times R_{EP} \times (1+R_{SP})^2}{R_{SP}} \right] B_s^{\frac{1}{6}} 
\label{ogvel}
\end{equation} 
The above empirical equation was derived using previous studies by \cite{Woo1977, Woo1978, Ho2002, Morabito2003, Yunqiu2015, BirdEdenhofer1990, Coles1989, Waldmeier1977}, and has been discussed in depth in \cite{Aggarwal2025a}.
Here, $f = 2.29$ GHz is the S-band signal frequency, $r$ is the closest distance of the radio signal path to the Sun, $R_{SP}$ and $R_{EP}$ are the Sun-probe and Earth-probe distances in astronomical units (AU), $R_{\odot}$ is the solar radius, and $B_s$ is the measured spectral width in Hz for a S band signal. The empirically derived constants $c_0 = 1.14 \times 10^{-24}$ and $k_0 = 1.687$ account for scaling and geometry, converting the measured spectral broadening into physically meaningful quantities \citep{Aggarwal2025a}. In simple terms, these equations tell us: the more the signal is broadened by the corona, the higher the electron density and/or the faster the solar wind. This method provides a direct way to estimate plasma properties without needing to assume a fixed turbulence model. It can also be adapted for other commonly used radio frequencies, such as X- and Ka-band, by including a simple wavelength scaling factor.

\subsection{Generalizing for other values of Turbulence Spectra}

The above relations assume a Kolmogorov-type turbulence spectrum with index $p = 11/3$ \citep{Aggarwal2025a, Aggarwal2025b}. However, the coronal plasma may exhibit a range of spectral indices depending on altitude and heliocentric distance. Following \citet{Morabito2009}, the spectral broadening scales with the observing wavelength $\lambda$ as:
\begin{equation}
    B \propto \lambda^{\frac{2}{p-2}},
\end{equation}
where $B$ is the measured broadening (in Hz) and $p$ is the spectral index of electron density fluctuations. For two different observing wavelengths, the ratio of broadening can be expressed as:
\begin{equation}
    \frac{B_s}{B_r} = \left( \frac{\lambda_s}{\lambda_r} \right)^{\frac{2}{p-2}}.
    \label{eq:lambda_scaling}
\end{equation}
Here, $B_s$ and $\lambda_s$ are the spectral width and the wavelength of the S band signal, while $B_r$ and $\lambda_r$ are the spectral width and the wavelength of the signal with arbitrary frequency $r$. Substituting $B_s$ from Equation \ref{eq:lambda_scaling} into the empirical relations above (eq. \ref{ogden} and \ref{ogvel}) yields a generalized form valid for different values of $p$ and wavelength ($\lambda_r$) used for radio occultation experiments:
\begin{equation}
    N_e = \frac{f}{r^2  \times  R_{SP}   \times r_{\odot}    (1+R_{SP})}
    \left( \frac{B_r}{c_0} \right)^{5/6}
    \left( \frac{13.2}{\lambda_r} \right)^{\frac{5}{3(p-2)}},
    \label{5}
\end{equation}
\begin{equation}
    v = k_0 \left[ \frac{r    R_{EP}    (1+R_{SP})^2}{R_{SP}} \right]
    B_r^{1/6} \left( \frac{13.2}{\lambda_r} \right)^{\frac{1}{3(p-2)}}.
    \label{6}
\end{equation}

\subsection{Turbulence Characterization and Plasma-Parameter Estimation from Frequency Fluctuations}

As the spacecraft radio signal passes close to the Sun, it encounters irregular fluctuations in electron density and velocity across a wide range of spatial and temporal scales, which perturb it. Measuring and analyzing these small frequency variations allows to infer properties of the turbulent medium through which the wave has propagated. The approach described here uses Doppler shift in the received signal to determine the slope of the coronal turbulence spectrum and estimate basic plasma parameters.

The observed Doppler data contain both predictable and random components. Deterministic effects caused by spacecraft motion, Earth’s rotation, and orbital geometry were modeled using the Krisher formula as given by \cite{Krisher1993}. The remaining signal contains the stochastic fluctuations produced mainly by the turbulent corona, along with slower instrumental and geometric drifts. To remove these slow trends while preserving the physical fluctuations of interest, each time series was detrended by fitting and subtracting a second-order polynomial. This approach filters out long-period variations that would distort the subsequent power spectrum and has been extensively discussed in \cite{Jain2022}.

\begin{figure*}
    \centering
    \includegraphics[width=0.9\linewidth]{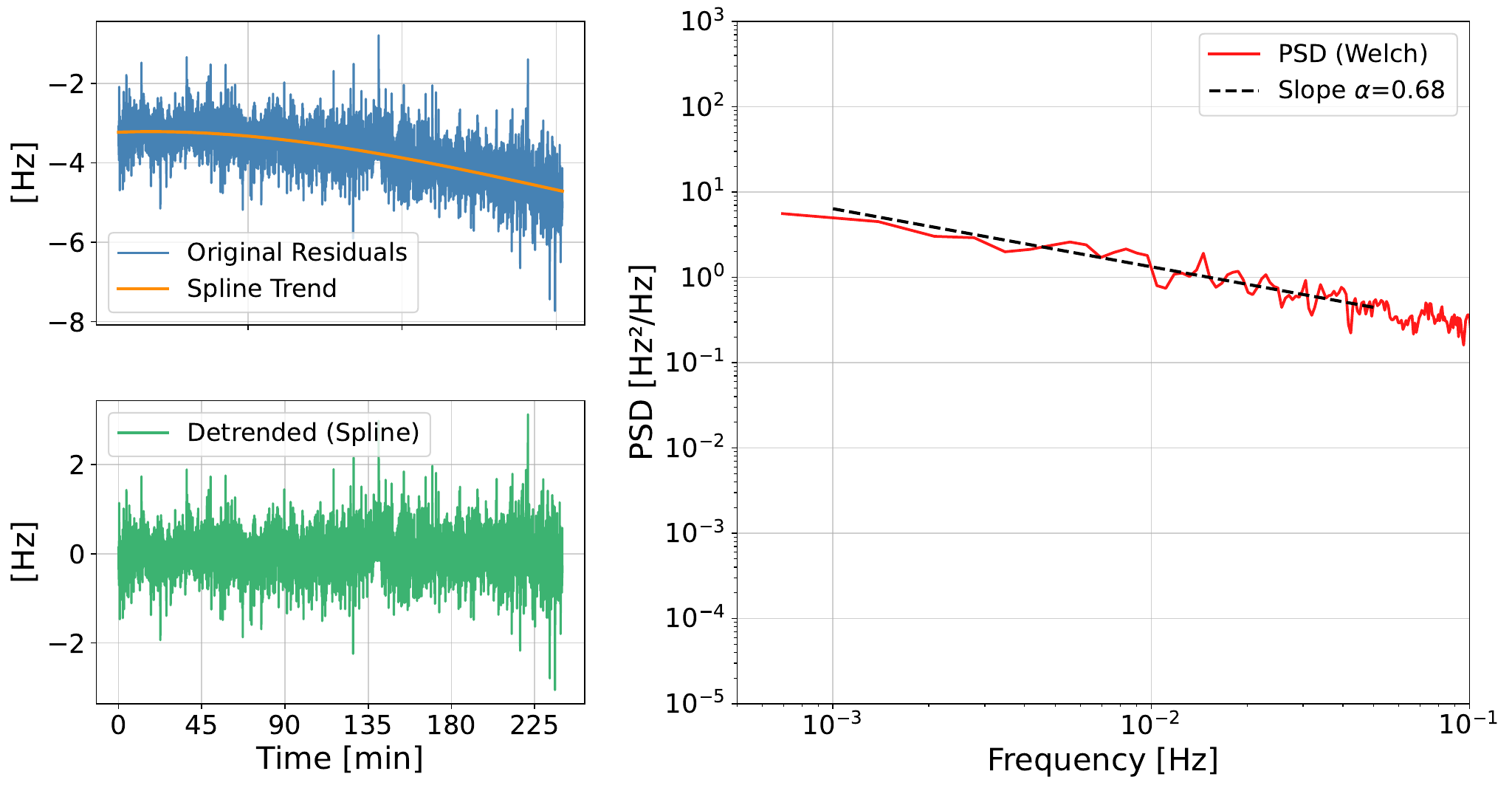}
    \caption{Top left: Time series of residuals in Hz are shown in blue, and the 2nd order polynomial fit used for detrending is shown in orange. Bottom left: Time series of residuals after detrending (Doppler FF). Right: PSD of the segment done using the Welch method. The black dashed line shows the slope-fit.}
    \label{fig:psd_plot}
\end{figure*}

\subsection{Power Spectral Density Estimation}

The detrended data were analyzed in the frequency domain using the Fourier transform to compute the power spectral density (PSD). To obtain stable PSD estimates, Welch's method was applied using the approach described in \cite{Yadav2025}. Figure \ref{fig:psd_plot} top left panel shows the time series of residuals in Hz in blue, and the 2nd order polynomial fit used for detrending in orange. The bottom left panel shows the time series of residuals after detrending (Doppler FF), and the right panel shows the PSD of the segment done using the Welch method with the black dashed line showing the slope fit. These segments were contiguous time windows of the detrended frequency time series used in Welch’s method for PSD estimation.

We employed a sliding window approach in which each dataset was segmented into windows of length equal to 10\% of the total dataset. Consecutive windows overlapped by 50\% of their length, meaning that each new window began halfway through the previous window. For example, if the dataset contains N samples, each window contains 0.1N samples, and successive windows starts 0.05N samples apart. This overlap preserves temporal continuity and reduces boundary effects while increasing the number of training segments. Further, rather than using fixed numerical window sizes, we adopted a percentage-based scheme to accommodate variations in dataset length. This ensured that the analysis remained scalable and consistent across datasets of differing sizes, improving the robustness and comparability of the results. This approach reduces variance in the PSD estimates and yields smooth, statistically reliable spectra. The usable frequency range for fitting was typically between $10^{-3}$ and $10^{-1}$\,Hz, avoiding the lowest frequencies dominated by residual drift and the highest frequencies where thermal and instrumental noise dominate \citep{Efimov2003, Efimov2005}.

In the inertial range of turbulent flow, the PSD generally follows a power-law form:

\begin{equation}
P(f) \propto f^{-\alpha},
\end{equation}
where $\alpha$ is the temporal spectral index, or the slope of the PSD of the frequency fluctuations. A straight-line fit to this relation in log-log space yields the best-fit $\alpha$. Figure \ref{fig:2016psd} shows PSDs of all days of experiment conducted using the  probe in the year 2016. It is clear from this Figure that as heliocentric distance increases, the overall amplitude of the fluctuations systematically decreases. Concurrently, the spectral shape transitions from a flattened low-frequency form at smaller solar offsets (upper curves near the horizontal blue dashed line for reference) to a steeper, negative power-law dependence characteristic of a Kolmogorov-type inertial subrange (lower curves near the red dashed reference line with slope $\alpha =-2/3$) at larger solar offsets. For isotropic turbulence, the slope of the PSD $\alpha$ are related to the spatial index by

\begin{equation}
p = \alpha + 3
\label{8}
\end{equation}
A Kolmogorov-type inertial cascade corresponds to $p = 11/3$, equivalent to $\alpha = 2/3$ under these definitions, where p denotes the spatial turbulence index.  
In the analyzed datasets, the derived slopes yield $p \approx 3.2$-$3.4$, indicating turbulence that is close to the Kolmogorov form but with small variations that may arise from anisotropy, compressibility, or dissipation effects which can be interpreted as signatures of changing turbulence regimes or of kinetic processes, such as wave-particle interactions, becoming significant in the near-Sun plasma.

\subsection{Derivation of Electron Density and Solar Wind Velocity}

Once the spectral slopes are determined, the physical plasma parameters can be estimated using empirical relations that link the observed signal broadening to the properties of the intervening medium. The measured spectral broadening (Doppler bandwidth) $B_r$ depends on the cumulative effect of electron density fluctuations along the path. Using equations \ref{5}, \ref{6} and \ref{8}, the electron density $N_e$ and solar wind velocity $v$ are given by
\begin{equation}
N_e = \frac{f}{r^2 R_{SP} r_{\odot}(1+R_{SP})} 
\left( \frac{B_r}{c_0} \right)^{5/6}
\left( \frac{13.2}{\lambda_r} \right)^{\frac{5}{3(\alpha + 1)}},
\end{equation}
\begin{equation}
v = k_0 \left[\frac{r R_{EP}(1+R_{SP})^2}{R_{SP}}\right]
B_r^{1/6}
\left( \frac{13.2}{\lambda_r} \right)^{\frac{1}{3(\alpha + 1)}},
\end{equation}
These equations relate the spectral characteristics of the received signal directly to the plasma conditions without assuming a fixed turbulence model. The use of the measured $\alpha$ makes the estimates sensitive to the actual turbulence conditions present during each observation.

The electron density and solar wind velocity profiles obtained from these radio occultation measurements have been compared with results from coronagraphs, white-light tomography, or in-situ measurements by missions such as Parker Solar Probe and Solar Orbiter \citep{Aggarwal2025a, Aggarwal2025b}. Agreement between these independent methods strengthens the case for using radio diagnostics as a quantitative probe of coronal turbulence. Deviations in spectral slopes from the canonical Kolmogorov form identify regions where energy dissipation or magnetic-field anisotropy becomes important, offering insight into how the solar wind is heated and accelerated near the Sun.

\section{Results}
\label{sec:results}

The data from the 2016 and 2022 Akatsuki radio occultation experiments were particularly valuable for this analysis as they sampled the solar wind under two distinct physical conditions -- slow and fast wind regimes spanning different phases of a solar cycle. The June 2016 conjunction occurred during the descending phase of Solar Cycle 24, a period of reduced solar activity characterized by a predominantly slow and structured solar wind originating near the streamer belt \citep{Chiba2022, Jain_2024}. In contrast, the October-November 2022 conjunction took place during the early rise of Solar Cycle 25, when equatorial coronal holes were well developed and produced sustained high-speed outflows \citep{Aggarwal2025b}.

The availability of X-band data from both events, recorded under well-calibrated geometry with the Akatsuki spacecraft and ground-based deep space antennas, enables direct evaluation of how the spectral properties of frequency fluctuations, electron density, and solar wind velocity vary with heliocentric distance and solar activity. Together, these datasets provide a unique opportunity to examine the evolution of coronal turbulence from 1.4 to 8.85 $R_{\odot}$ under differing magnetic and dynamic boundary conditions, offering complementary perspectives on the development of the inertial cascade and its role in solar wind acceleration. 

\subsection{Turbulence spectral characteristics during the 2016 conjunction}
\label{sec:2016}

\begin{figure*}
    \centering
    \includegraphics[width=\linewidth]{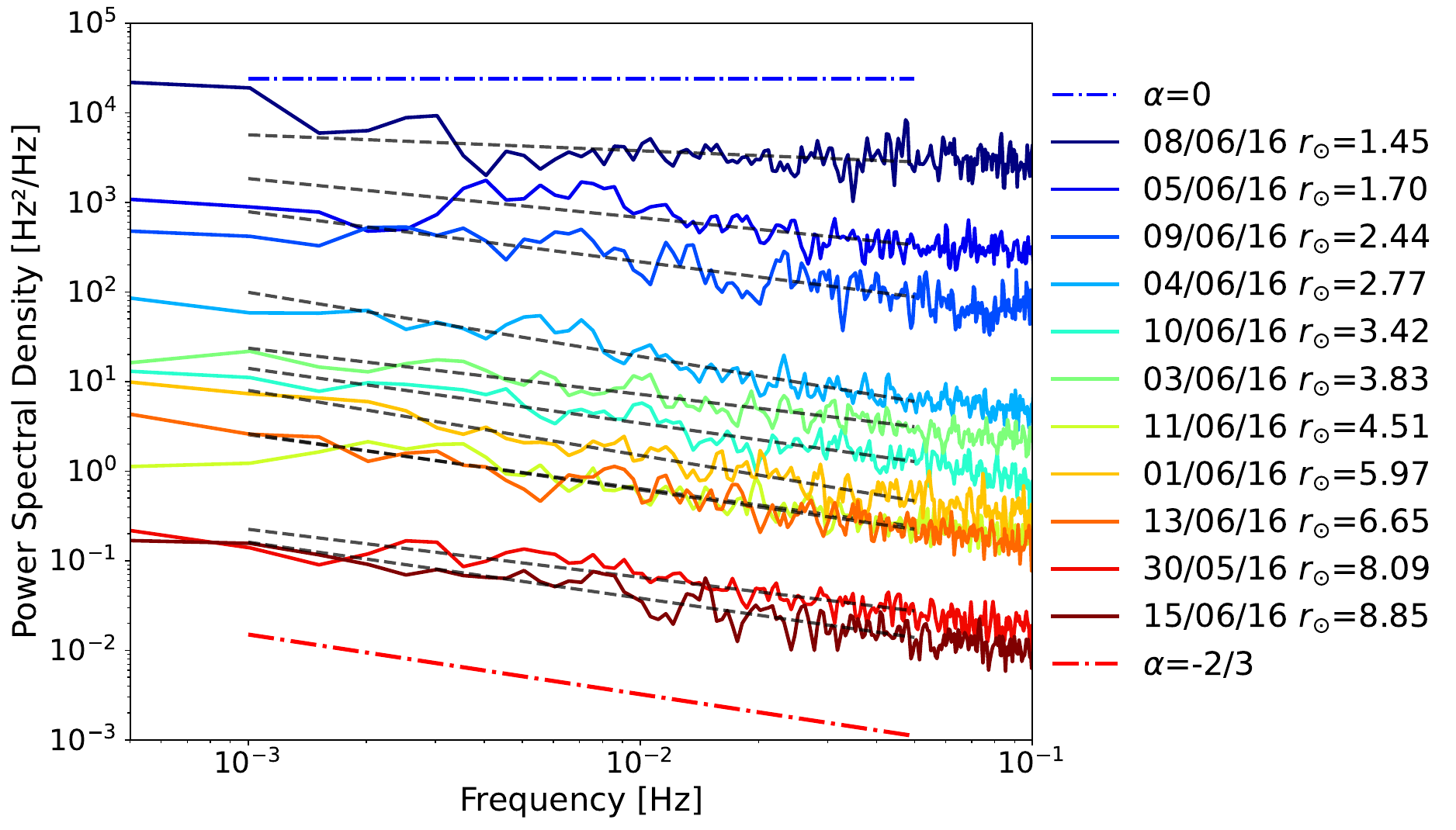}
    \caption{The PSDs of all days of experiment conducted using the  probe in the year 2016. The corresponding legend marker also shows the heliocentric distance for each experiment.}
    \label{fig:2016psd}
\end{figure*}

As shown in Figure \ref{fig:2016psd}, the temporal power spectral indices derived from the X-band radio signal fluctuations during the 2016 Venus solar conjunction reveal a clear radial evolution in the turbulence spectrum between 1.45 and 8.85 $R_{\odot}$. The results, summarized in Table \ref{tab:slope2016}, show that the spectral amplitude slopes ($\alpha$) vary between 0.18 and 0.73, corresponding to spatial turbulence indices ($p = \alpha + 3$) ranging from 3.18 to 3.73.

The lowest $p$ value of 3.18 occurs near the closest solar approach ($1.45\, R_{\odot}$), indicating enhanced low-frequency power and a relative flattening of the spectrum. This behavior is characteristic of regions where large-scale magnetic structures or streamer boundaries dominate the dynamics and inject energy into the coronal plasma \cite{Imamura2005, Wexler2016}. Between 2 and 3 $R_{\odot}$, the spectral index steepens rapidly toward $p \approx 3.6$, marking the onset of a developed turbulent cascade. Beyond $\sim3 \, R_{\odot}$, $p$ remains stable at $3.6$-$3.7$, consistent with the canonical Kolmogorov scaling ($p = 11/3$). This stabilization indicates the transition to a quasi-stationary, fully developed magnetohydrodynamic (MHD) turbulence regime where nonlinear interactions dominate over direct energy injection.
\begin{table*}
\centering
\begin{tabular}{lcccccccccc}
\hline
Date & R ($R_{\odot}$) & Slope ($\alpha$) & p &
\multicolumn{2}{c}{Electron Density (m$^{-3}$)} &
\multicolumn{4}{c}{Velocity (km s$^{-1}$)} &
Diff (\%)\\
\cline{5-6}\cline{7-10}
 &  &  &  & Ref. & This Study & Ref. & Err & This Study & Err &  \\
\hline
30 May & 8.09 & 0.54 & 3.54 & 9.97E+10 & 8.76E+10 & 138.19 & 1.13 & 140.41 & 0.73 & 1.58\\
1 Jun & 5.97 & 0.73 & 3.73 & 1.61E+11 & 1.60E+11 & 105.84 & 0.92 & 103.55 & 0.64 & -2.21\\
3 Jun & 3.83 & 0.52 & 3.52 & 3.36E+11 & 5.91E+11 & 72.26 & 0.60 & 72.36 & 0.46 & 0.13\\
4 Jun & 2.77 & 0.72 & 3.72 & 6.01E+11 & 1.19E+12 & 55.08 & 0.34 & 52.65 & 0.25 & -4.62\\
5 Jun & 1.70 & 0.44 & 3.44 & 1.12E+12 & 4.54E+12 & 35.58 & 0.18 & 34.81 & 0.13 & -2.21\\
8 Jun & 1.45 & 0.18 & 3.18 & 1.31E+12 & 8.69E+12 & 30.72 & 0.15 & 31.88 & 0.12 & 3.64\\
9 Jun & 2.44 & 0.56 & 3.56 & 7.66E+11 & 1.92E+12 & 49.46 & 0.27 & 48.67 & 0.20 & -1.62\\
10 Jun & 3.42 & 0.61 & 3.61 & 3.56E+11 & 6.54E+11 & 64.28 & 0.66 & 63.02 & 0.53 & -1.99\\
11 Jun & 4.51 & 0.60 & 3.60 & 2.35E+11 & 3.37E+11 & 81.90 & 0.98 & 81.17 & 0.78 & -0.91\\
13 Jun & 6.65 & 0.63 & 3.63 & 1.60E+11 & 1.51E+11 & 118.96 & 1.09 & 119.03 & 0.82 & 0.05\\
15 Jun & 8.85 & 0.63 & 3.63 & 1.07E+11 & 7.71E+10 & 155.07 & 1.61 & 155.16 & 1.21 & 0.05\\
\hline
\end{tabular}
\caption{Derived parameters from 2016  radio occultation data compared with results from \citet{Aggarwal2025b}.}
\label{tab:slope2016}
\end{table*}

The derived plasma parameters support this interpretation. Electron densities decrease monotonically from $8.69\times10^{12} \, \mathrm{m^{-3}}$ at $1.45\, R_{\odot}$ to $7.7\times10^{10}\, \mathrm{m^{-3}}$ at $8.85 \, R_{\odot}$, in close agreement with empirical coronal models \citep[e.g.,][]{Edenhofer1977, Leblanc1998, Wexler2019}. Over the same interval, the solar wind velocity increases from $\sim 31.88\, \mathrm{km s^{-1}}$ to $155 \, \mathrm{km s^{-1}}$, reflecting the expected acceleration of the slow solar wind through the inner corona. The differences between the present results and those of \citet{Aggarwal2025b} remain within 5\% for all dates, validating the turbulence-based spectral derivation method. It is also important to note that these results show slight deviation from the results discussed by \cite{Chiba2022} using the same dataset, meaning that the conditions are not isotropic for slow solar wind speeds. 

Overall, the 2016  radio occultation data indicate a gradual transition from energy-injection-dominated turbulence near the solar surface to a developed Kolmogorov-like inertial range beyond 3 $R_{\odot}$. The observed flattening below this height likely reflects magnetically structured slow-wind regions with enhanced compressibility and energy input from the lower corona.

\subsection{Turbulence spectral characteristics during the 2022 coronal-hole occultation}
\label{sec:2022}

\begin{figure*}
    \centering
    \includegraphics[width=\linewidth]{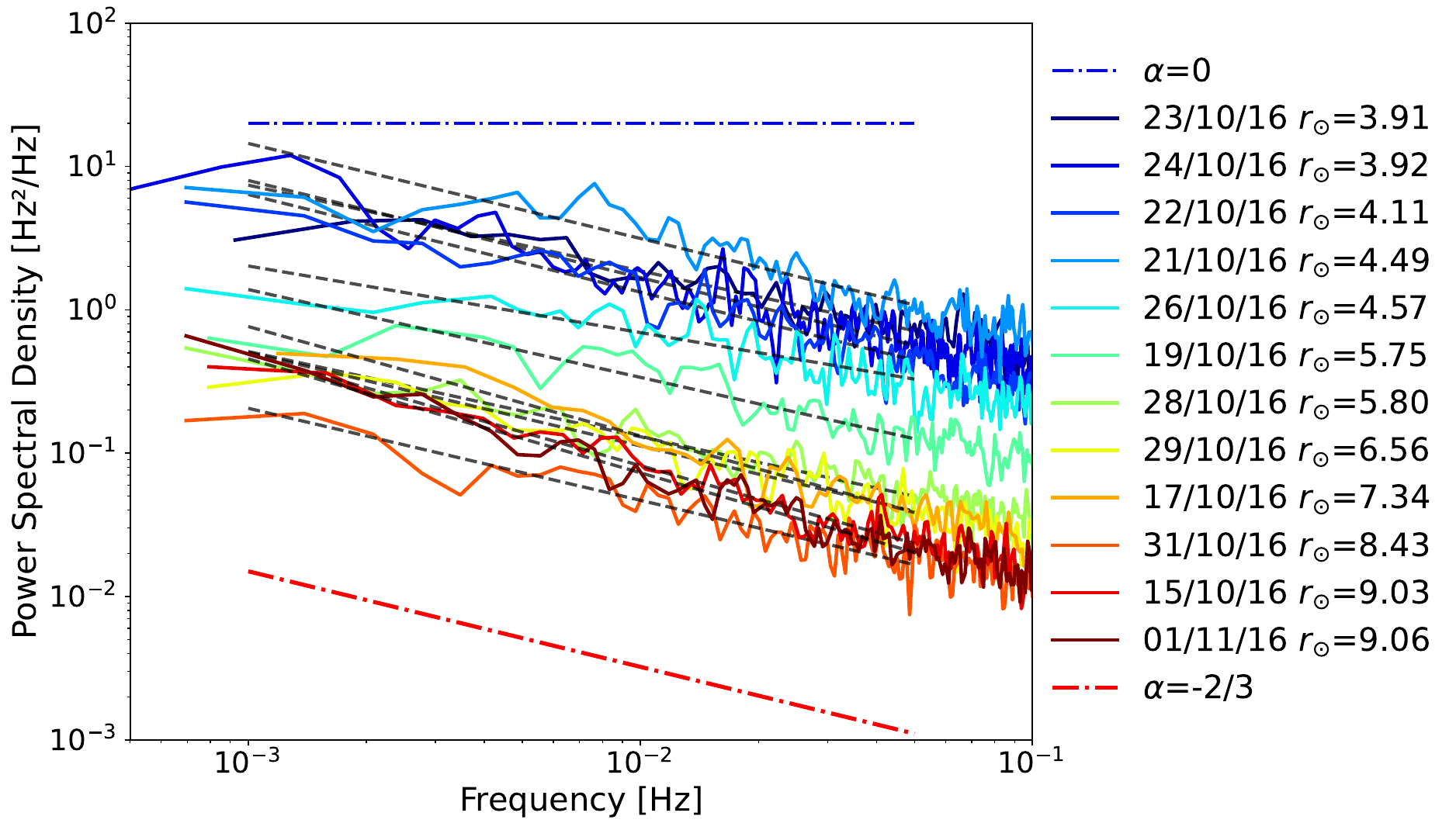}
    \caption{Same as Figure \ref{fig:2016psd}, but for the year 2022.}
    \label{fig:2022psd}
\end{figure*}

The October-November 2022 Venus solar conjunction provided an opportunity to probe the high-speed solar wind emerging from equatorial coronal holes. Figure \ref{fig:2022psd} shows the PSDs  of all days of experiment conducted in the year 2022. The derived turbulence spectral slopes ($\alpha$) range from 0.46 to 0.79, corresponding to spatial indices ($p = 3.46$-$3.79$). Though a bit noisy compared to 2016, these values remain tightly clustered around the Kolmogorov index ($p = 11/3$), showing variation within expectations with heliocentric distance (Table \ref{tab:slope2022}).

Unlike the 2016 slow-wind case, in 2022, the slopes are much steeper even for the lower radial distances. The near-constant $p$ values from $3.9$ to $9.1\, R_{\odot}$ indicate that the turbulence in coronal-hole outflows is homogeneous and already well developed close to the solar surface. The absence of low-$p$ excursions suggests limited influence from large-scale magnetic structures or transient activity along the line of sight, consistent with the stable, persistent coronal-hole conditions during the observation period.
\begin{table*}
\centering
\begin{tabular}{lcccccccccc}
\hline
Date & R ($R_{\odot}$) & Slope ($\alpha$) & p &
\multicolumn{2}{c}{Electron Density (m$^{-3}$)} &
\multicolumn{4}{c}{Velocity (km s$^{-1}$)} &
Diff (\%)\\
\cline{5-6}\cline{7-10}
 &  &  &  & Ref. & This Study & Ref. & Err & This Study & Err &  \\
\hline
15 Oct & 9.03 & 0.78 & 3.78 & 3.47E+10 & 2.58E+10 & 352.36 & 1.98 & 357.12 & 0.49 & 1.33\\
17 Oct & 7.34 & 0.76 & 3.76 & 4.58E+10 & 4.10E+10 & 287.43 & 2.77 & 294.81 & 1.01 & 2.50\\
19 Oct & 5.75 & 0.61 & 3.61 & 5.94E+10 & 7.42E+10 & 226.17 & 2.31 & 237.23 & 0.88 & 4.66\\
21 Oct & 4.49 & 0.66 & 3.66 & 1.08E+11 & 1.69E+11 & 188.09 & 2.90 & 194.62 & 1.28 & 3.35\\
22 Oct & 4.11 & 0.68 & 3.68 & 1.03E+11 & 1.71E+11 & 168.93 & 2.63 & 173.75 & 1.14 & 2.77\\
23 Oct & 3.91 & 0.60 & 3.60 & 1.09E+11 & 2.02E+11 & 161.75 & 2.42 & 167.74 & 1.05 & 3.57\\
24 Oct & 3.92 & 0.68 & 3.68 & 1.31E+11 & 2.22E+11 & 169.42 & 2.96 & 171.07 & 1.30 & 0.96\\
26 Oct & 4.57 & 0.46 & 3.46 & 8.76E+10 & 1.56E+11 & 193.05 & 3.08 & 199.19 & 1.33 & 3.08\\
28 Oct & 5.80 & 0.59 & 3.59 & 5.96E+10 & 7.80E+10 & 240.16 & 3.08 & 240.53 & 1.20 & 0.15\\
29 Oct & 6.56 & 0.65 & 3.65 & 5.05E+10 & 5.63E+10 & 269.75 & 2.49 & 267.32 & 0.86 & -0.91\\
31 Oct & 8.43 & 0.64 & 3.64 & 3.86E+10 & 3.40E+10 & 336.08 & 2.97 & 342.64 & 1.03 & 1.91\\
1 Nov & 9.06 & 0.79 & 3.79 & 3.80E+10 & 2.73E+10 & 370.38 & 3.53 & 363.36 & 1.24 & -1.93\\
\hline
\end{tabular}
\caption{Derived parameters from 2022 radio occultation data compared with results from \citet{Aggarwal2025b}.}
\label{tab:slope2022}
\end{table*}

Electron densities derived for the 2022 event range from $2.02\times10^{11} \mathrm{m^{-3}}$ at $3.9 \, R_{\odot}$ to $2.58\times10^{10}\, \mathrm{m^{-3}}$ at $9.1 \,R_{\odot}$, decreasing systematically with radial distance \citep{Doschek1997, Gallagher1999}. Over the same interval, solar wind velocities increase from $171 \,\mathrm{km s^{-1}}$ to $363 \, \mathrm{km s^{-1}}$, marking the transition from the low corona to the fully accelerated fast-wind regime. These trends agree with the study by \cite{Jain2024} in which they applied the isotropic turbulence method, showing that fast winds indeed show behavious close to isotropic, and retains an Alfv\'enic, low-compressibility character.

\begin{figure}
    \centering
    \includegraphics[width=\linewidth]{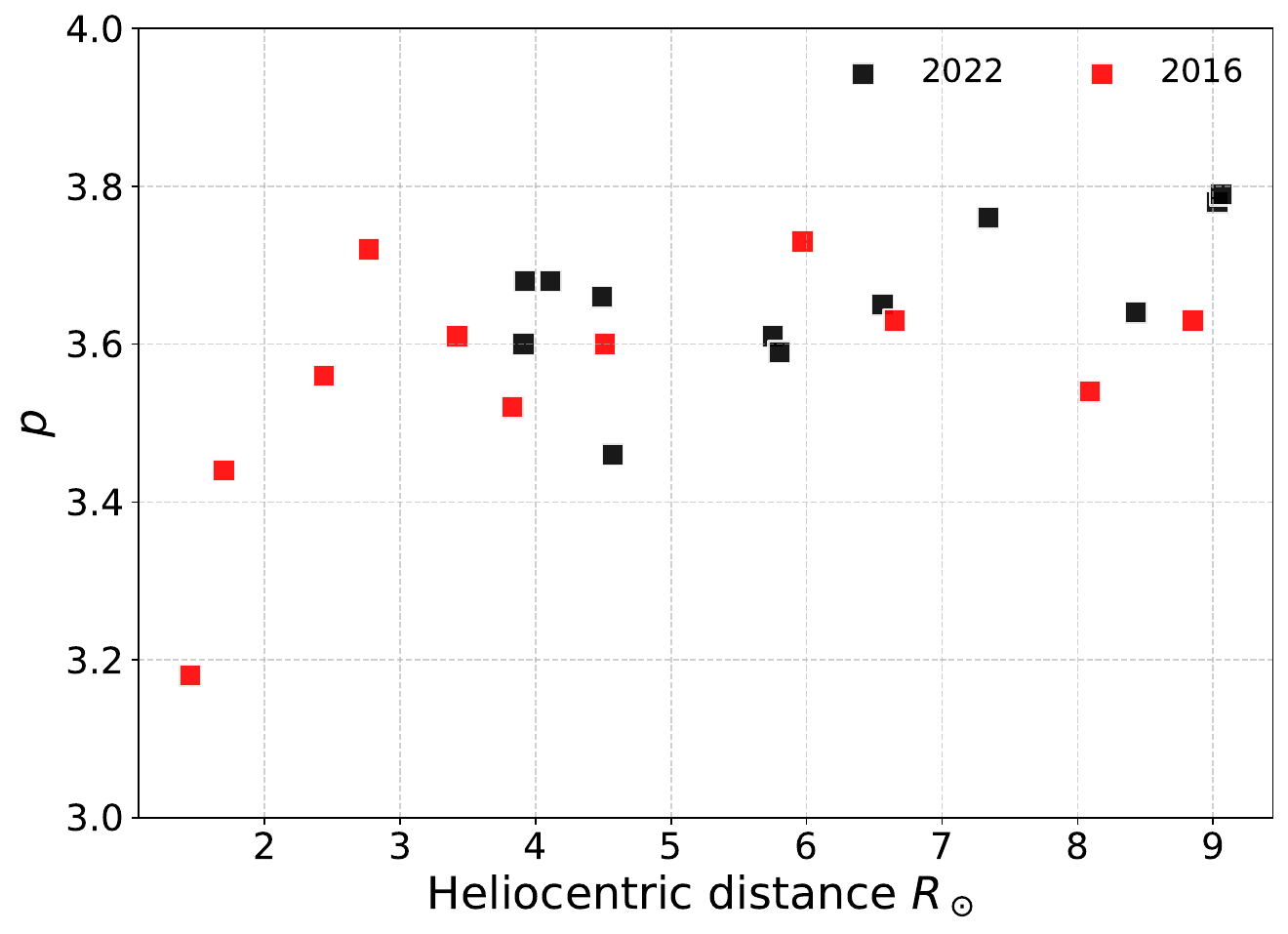}
    \caption{Values of $p$ as a function of radial distance. The values of the quantities are shown in red for 2016, and in black for 2022.}
    \label{fig:kolmog}
\end{figure}

A mild steepening of the spectrum ($p\sim3.78$-$3.79$) is seen at the largest heliocentric distances ($>7 \, R_{\odot}$), possibly indicating enhanced dissipation or a shift toward anisotropic Alfv\'enic turbulence as the solar wind becomes collisionless. Conversely, a modest flattening around $4.5 \, R_{\odot}$ ($p=3.46$) may reflect local magnetic reconnection or small-scale energy input near streamer-hole interfaces. The close agreement between these results and those reported by \citet{Aggarwal2025b} (differences within 5\%) further validates the turbulence-based derivation Figure \ref{fig:kolmog} shows the values of $\alpha$ and $p$ as a function of radial distance, in red for 2016, and in black for 2022.

Taken together, the 2022 observations demonstrate that turbulence in coronal-hole outflows follows a near-Kolmogorov spectrum throughout 3-9\, $R_{\odot}$, consistent with a mature, Alfv\'enic regime. In contrast to the 2016 slow-wind case, where $p$ varied strongly with distance, the 2022 data indicate an already well-developed cascade, supporting the interpretation that fast-wind turbulence forms lower in the corona and evolves more steadily with heliocentric distance.

\section{Discussion and Concluding remarks}
\label{sec:conclusions}
\begin{figure}
    \includegraphics[width=1\linewidth]{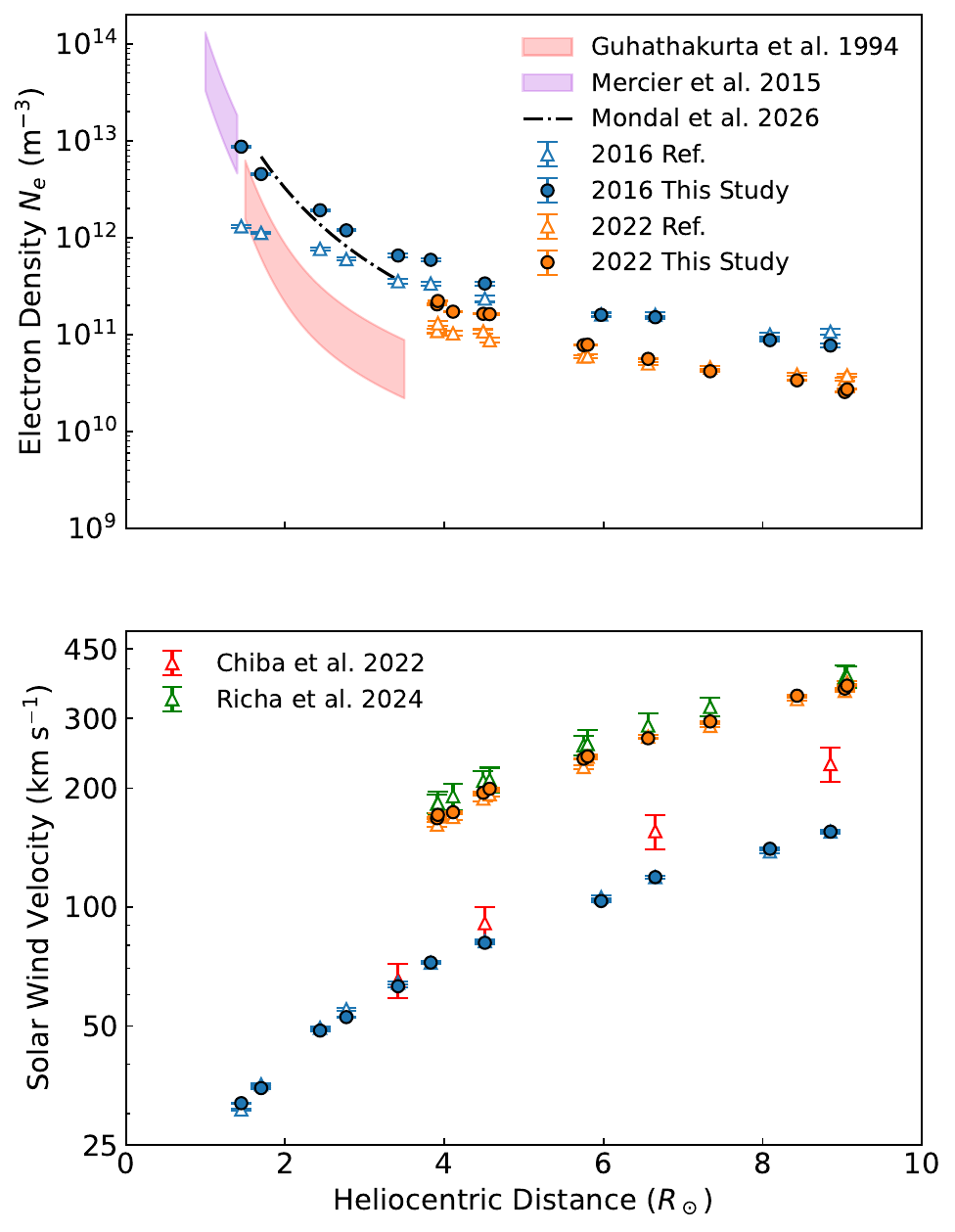}
    \caption{Comparison of electron density and solar wind velocity derived from Akatsuki radio occultation experiments during the 2016 and 2022 superior conjunctions. Blue symbols denote the 2016 slow-wind event, orange symbols the 2022 fast-wind event. Triangles represent reference values from \citet{Aggarwal2025b}, while circles show the present turbulence-based results. Both datasets exhibit the expected decrease in $N_e$ and increase in $v$ with heliocentric distance. The results for solar wind speeds by \citet{Chiba2022, Jain2024} are shown in red and green triangles respectively.}
    \label{fig:velden_comp}
\end{figure}

The derived plasma parameters for both the 2016 and 2022  occultation campaigns are summarized in Figure \ref{fig:velden_comp}. While the points in blue and orange show the comparison between our results here and those in our previous study \citep{Aggarwal2025b}, we also compare our velocity measurements with those of \cite{Chiba2022, Jain2024} in red and green triangles. Comparisons with models for electron density like \citet{Guhathakurta1994, Mercier2015, Mondal2026} show that the measurements made in this study are quite close as well.
For the 2016 conjunction, which sampled streamer-belt regions dominated by the slow solar wind, both the reference and derived values show electron densities decreasing from $\sim10^{12} \mathrm{m^{-3}}$ near $1.5 \, R_{\odot}$ to $\sim10^{11} \mathrm{m^{-3}}$ beyond $8\, R_{\odot}$, accompanied by gradual acceleration of the plasma from about $30$ to $150\, \mathrm{km s^{-1}}$, consistent with streamer-belt density models in this radial range. It is important to note that while the measurements follow the general trend, they do deviate slightly, which makes sense due to the assumption of isotropic density spectrum \citep{Chiba2022,Chiba2023}. In contrast, the 2022 event, which probed equatorial coronal hole outflows associated with the fast solar wind, exhibits lower overall densities and systematically higher velocities, increasing from $\sim \,170$ to over $350 \, \mathrm{km s^{-1}}$ between $4$ and $9\, R_{\odot}$.

The close correspondence between our results and those obtained in \citet{Aggarwal2025b}, with differences typically below 5\%, reinforces the reliability of the method based on spectral broadening and turbulence analysis. The results from 2016 and 2022 also show the same behavior as demonstrated by \cite{Yadav2025}, and show that while for the quiet periods anisotropic turbulence is seen, the extreme events display quasi-isotropic behavior. These results confirm that radio occultation can recover realistic radial trends in both $N_e$ and $v$ that are consistent with independent coronal models and in situ solar wind measurements.

The two datasets together capture the transition from the dense, magnetically structured slow-wind corona to the more rarefied, Alfv\'enic fast-wind environment, providing observational constraints on density fluctuations, velocity gradients, and turbulence evolution across the inner heliosphere.

As we note, the Akatsuki radio occultation experiments demonstrate that spectral broadening analysis of spacecraft signals can quantitatively characterize coronal turbulence and infer key plasma parameters across heliocentric distances of $1.4-9\, R_{\odot}$. By analyzing the frequency-domain properties of Doppler residuals, we establish a direct link between radio signal fluctuations and the turbulent density and velocity structure of the solar corona.
The derived spectral slopes ($\alpha = 0.18$-$0.79$) and corresponding spatial indices ($p = 3.18$-$3.79$) trace the evolution of turbulence from the magnetically structured, energy-injection-dominated inner corona to the mature, Kolmogorov-like inertial regime of the outer corona. The 2016 slow-wind conjunction reveals a progressive steepening of the turbulence spectrum beyond $\sim3\, R_{\odot}$, marking the onset of a well-developed cascade, whereas the 2022 coronal-hole occultation shows slopes expected through the region of 3-9 $R_{\odot}$, indicative of homogeneous, Alfv\'enic turbulence within fast-wind outflows. This distinction reflects the different stages of turbulence evolution in slow and fast solar wind sources.

The electron densities derived from the spectral broadening analysis span $8.7\times10^{12}$ to $7.7\times10^{10}$ m$^{-3}$ between 1.45 and 9 $R_{\odot}$, while solar wind velocities increase from $\sim30$ to $360$ km s$^{-1}$ over the same range. Both trends are consistent with classical coronal expansion models and in-situ measurements by the Parker Solar Probe. Comparisons with other models like \citet{Guhathakurta1994, Mercier2015, Mondal2026} show that the estimates are quite close as well. The close agreement (within 5\%) between these results and those obtained using independent assumptions of Kolmogorov turbulence confirms the robustness of the frequency-fluctuation-based diagnostic approach.

This work improves the earlier analysis by replacing fixed-spectrum assumptions with direct turbulence-slope measurements from the spectral broadening \citep{Aggarwal2025b}. Unlike direct spacecraft sampling, which is limited by orbital constraints, radio occultation samples a broad range of heliocentric distances and latitudes during conjunction geometries, providing a statistically rich dataset for turbulence characterization. The methodology developed here can be readily extended to multi-frequency occultations and cross-instrument analyses with ongoing missions such as Parker Solar Probe and Solar Orbiter, enabling cross-validation of remote and in-situ turbulence measurements.

The derived spectral indices and plasma parameters also reveal how the nature of coronal turbulence changes with distance from the Sun. The derived spectral slopes trace two distinct behaviours: a transition from sub-Kolmogorov flattening to a developed cascade in 2016, and an already near-Kolmogorov spectrum across all sampled distances in 2022. The corresponding spatial indices suggest that the turbulence in the fast-wind intervals remains roughly isotropic up to several solar radii, whereas the slow-wind intervals exhibit measurable deviations from isotropy, though localized deviations may occur where compressive or anisotropic structures dominate. The measured spectral broadening and the fitted temporal index $\alpha$ incorporate LOS-averaged effects, though residual anisotropy can introduce modest deviations from isotropic estimates. The conversion from these LOS measurements to perpendicular density and velocity uses formulas based on isotropic turbulence, where the amplitude of the fluctuations is the same in every direction.

In the corona, however, this is not always true, because the fluctuations are organised relative to the local magnetic-field direction: they are typically larger in the direction perpendicular to the local field and smaller in the direction parallel to it. When this difference is present, the LOS broadening no longer follows the isotropic relation exactly, and the density and velocity obtained from it can shift slightly above or below their true perpendicular values. For the datasets analysed here, the derived densities and velocities agree with independent estimates from the same observations, showing that the level of this directional difference is not large enough to affect the results in a significant way. A natural extension of this work would be to incorporate anisotropy, which would refine the interpretation without altering the turbulence-spectrum analysis used in this study.

Future radio science experiments with higher cadence and multi-band capabilities, particularly those anticipated with the Square Kilometre Array (SKA) and next-generation deep-space missions, will allow detailed mapping of electron density fluctuations, spectral anisotropies, and the spatial variation of the turbulence index. Such measurements will bridge the remaining gap between remote-sensing diagnostics and direct in-situ plasma measurements, offering a unified view of energy transport, wave-particle interactions, and solar wind acceleration in the inner heliosphere.

\section*{Data Availability}
Solar occultation and SPICE Kernel data from Akatsuki probe obtained from the Akatsuki Radio Science (RS) Data Archive at \href{https://darts.isas.jaxa.jp/planet/project/VCO/index}{link} was used for the experiment. 

\section*{Acknowledgements}
We would like to thank the Akatsuki mission team for monitoring the Akatsuki radio signals. K.A. sincerely thanks Head, HRDD, VSSC, ISRO and Director, SPL for the opportunity to visit SPL as part of this project. We would like to thank the Akatsuki mission team for monitoring the Akatsuki radio signals. Help provided by R. S. Simi in providing the data is also gratefully acknowledged. The Prime Minister's Research scholarship (PMRF) program, Ministry of Education, Government of India, awarded K.A. a research scholarship (PMRF-2103356). K.A. and A.D. acknowledge the use of facilities procured through the funding via the Department of Science and Technology, Government of India sponsored DST-FIST grant no. SR/FST/PSII/2021/162 (C) awarded to the DAASE, IIT Indore.

\bibliographystyle{mnras}
\bibliography{biblio_turbulence} % if your bibtex file is called example.bib
\bsp	% typesetting comment
\label{lastpage}
\end{document}